\documentclass[12pt,epsf]{article}
\usepackage{graphicx}
\usepackage{amsmath}
\begin{document}

\def\a{\alpha}
\def\b{\beta}
\def\c{\varepsilon}
\def\d{\delta}
\def\e{\epsilon}
\def\f{\phi}
\def\g{\gamma}
\def\h{\theta}
\def\k{\kappa}
\def\l{\lambda}
\def\m{\mu}
\def\n{\nu}
\def\p{\psi}
\def\q{\partial}
\def\r{\rho}
\def\s{\sigma}
\def\t{\tau}
\def\u{\upsilon}
\def\v{\varphi}
\def\w{\omega}
\def\x{\xi}
\def\y{\eta}
\def\z{\zeta}
\def\D{\Delta}
\def\G{\Gamma}
\def\H{\Theta}
\def\L{\Lambda}
\def\F{\Phi}
\def\P{\Psi}
\def\S{\Sigma}

\def\o{\over}
\def\beq{\begin{eqnarray}}
\def\eeq{\end{eqnarray}}
\newcommand{\wt}{\widetilde}
\newcommand{\gsim}{ \mathop{}_{\textstyle \sim}^{\textstyle >} }
\newcommand{\lsim}{ \mathop{}_{\textstyle \sim}^{\textstyle <} }
\newcommand{\vev}[1]{ \left\langle {#1} \right\rangle }
\newcommand{\bra}[1]{ \langle {#1} | }
\newcommand{\ket}[1]{ | {#1} \rangle }
\newcommand{\EV}{ {\rm eV} }
\newcommand{\KEV}{ {\rm keV} }
\newcommand{\MEV}{ {\rm MeV} }
\newcommand{\GEV}{ {\rm GeV} }
\newcommand{\TEV}{ {\rm TeV} }
\def\diag{\mathop{\rm diag}\nolimits}
\def\Spin{\mathop{\rm Spin}}
\def\SO{\mathop{\rm SO}}
\def\O{\mathop{\rm O}}
\def\SU{\mathop{\rm SU}}
\def\U{\mathop{\rm U}}
\def\Sp{\mathop{\rm Sp}}
\def\SL{\mathop{\rm SL}}
\def\tr{\mathop{\rm tr}}

\def\IJMP{Int.~J.~Mod.~Phys. }
\def\MPL{Mod.~Phys.~Lett. }
\def\NP{Nucl.~Phys. }
\def\PL{Phys.~Lett. }
\def\PR{Phys.~Rev. }
\def\PRL{Phys.~Rev.~Lett. }
\def\PTP{Prog.~Theor.~Phys. }
\def\ZP{Z.~Phys. }


\baselineskip 0.7cm

\begin{titlepage}

\begin{flushright}
IPMU10-0139
\end{flushright}

\vskip 1.35cm
\begin{center}
{\large \bf
A CP-safe solution of the $\mu$/$B\mu$ problem of gauge mediation
}
\vskip 1.2cm
Jason L. Evans, Matthew Sudano, and  Tsutomu T. Yanagida
\vskip 0.4cm

{\it  Institute for the Physics and Mathematics of the Universe, \\
University of Tokyo, Chiba 277-8583, Japan}

\vskip 1.5cm

\abstract{We construct a model that naturally generates $\mu$ and $B$ of the same order without producing large CP violating phases.  This is easily accomplished once one permits these mass scales to be determined independently of the ordinary gauge-mediated soft masses.  The alignment of phases is shown to emerge dynamically upon coupling to supergravity and is not unique to the model presented here.}

\end{center}
\end{titlepage}

\setcounter{page}{2}

\section{Introduction}

Gauge mediation \cite{GM} is a well-motivated mechanism for communicating supersymmetry (SUSY) breaking. It is a simple and calculable scenario that naturally solves the flavor-changing neutral current problem. However, gauge mediation is not without problems. One of the main difficulties is in explaining the magnitudes of the Higgs mass parameters, $\mu$ and $B\mu$,
\begin{equation}
V\supset|\mu|^2(H_u^\dagger H_u+H_d^\dagger H_d)+(B\mu H_uH_d+h.c.).
\end{equation}
To accomplish electroweak symmetry breaking without considerable fine tuning, the SUSY preserving mass, $\mu$, and the soft-breaking parameter, $B$, should each be near the weak scale.  However, the simplest attempts at generating these scales in gauge mediation lead to $|B/\mu|\sim16\pi^2$ \cite{Bmu}.  This is the case, for example, in ordinary gauge mediation if one employs the same singlet spurion to generate all of the MSSM mass scales.

In attempting to solve this problem, one is led to introduce additional fields and couplings, but this can lead to a new problem.  In introducing new complex parameters, one often badly violates CP\footnote{Because the $A$ terms in gauge mediation are small, we can neglect the CP violation in these terms.}.  In particular, when the gaugino masses and $B\mu$ are generated by independent mechanisms, one cannot in general use the $R$-symmetry to eliminate the dangerous phases.

In this short paper, we propose a simple model that solves the above problems (see \cite{musol} for some alternate approaches).  We allow the Higgses to couple to a singlet SUSY-breaking field that naturally gives $|\mu|\sim |B|$ at tree-level.  The ordinary messengers of gauge mediation are given mass by a different field, but we show that the relevant phases, which appear to be free in the rigid theory, are in fact highly constrained by supergravity dynamics.  In fact, we will see that supergravity is not an essential ingredient; our phases would also be fixed by corrections to the K\"ahler potential, and in general appropriately chosen interactions can do the job.  However, since the cosmological constant is an essential ingredient in any realistic theory, we focus of the former mechanism.

\section{A gauge mediation model}

The messenger sector considered here is quite simple. We take a singlet superfield $S$ and a pair of messengers
$\Psi$ and ${\wt\Psi}$ which respectively transform as a ${\bf 5}$ and ${\bf 5^*}$ of $SU(5)_{GUT}$. The superpotential is
\begin{equation}\label{wmess}
W= \frac{\l}{3} S^3 + kS\Psi{\wt \Psi}.
\end{equation}
We use $S$ to choose $\l$ real and negative for future convenience.
Now we assume that supersymmetry is broken in some other sector, inducing a tachyonic SUSY-breaking soft mass for the scalar component of $S$ (this is the case in \cite{IFY}, which we review in the appendix). Once supersymmetry is broken, the scalar potential is given by
\begin{equation}
V= -m^2_S|S|^2 + |\l S^2+k\Psi\wt\Psi|^2+|kS\Psi|^2+|kS\wt\Psi|^2.
\end{equation}
This gives $\vev\Psi=0=\langle\wt\Psi\rangle$ and
\begin{equation}\label{svev}
\vev S=e^{i\d_S}\frac{m_S}{\sqrt{2}\l},\qquad F_S =- \l \langle S^\dagger\rangle^2,
\end{equation}
where $\d_S$ is an undetermined phase.  The one-loop gaugino masses then follow from (\ref{svev});
\begin{equation}\label{gmass}
m_i=-\frac{\alpha_i}{4\pi}\l |S|e^{-i3\d_S} g(x),
\end{equation}
where $x=\left|\l/k\right|$ and $g(x)$ is defined in \cite{GGM}. We see that the phase of the gaugino masses is determined by the phase of $S$.

We could use the $R$-symmetry to rotate away this phase, but we will instead use this freedom to choose the gravitino mass, $m_{3/2}$, real and positive.  Upon coupling to supergravity this parameter appears in the superpotential, $W\supset m_{3/2}M_P^2$, where $M_P=2.4\times 10^{18}$GeV is the Planck scale.  This term explicitly breaks the $R$-symmetry and leads to the dynamical determination of the phase of $S$.

To see this, consider the scalar potential for our model \cite{WesABag},
\begin{equation}
V= e^{K/M_P^2}\left(g^{\bar SS}\left|W_S +\frac{K_S W}{M_P^2}\right|^2 -3\frac{ |W|^2}{M_P^2}\right)+V_{soft}.
\end{equation}
where
\begin{equation}
W=\frac\l3 S^3 + kS\Psi{\bar \Psi}+m_{3/2}M_P^2, \qquad K=|S|^2-\frac h{\Lambda^2}|S|^4 +...,
\qquad g_{i\bar j} =\partial_i\partial_{\bar j}K,
\end{equation}
$V_{soft}$ contains our SUSY breaking mass, $h$ is a real constant, and $\Lambda$ is the cutoff scale.  Writing $S=|S|e^{i\d_S}$ the relevant terms in the potential are
\begin{equation}
V\supset-m_S^2|S|^2+|\l|^2 |S|^4+4\l m_{3/2}\frac{h}{\Lambda^2}|S|^5\cos3\d_S. \label{SV}
\end{equation}
Assuming that $h$ is positive\footnote{If one takes $h$ negative, $\d_S$ is simply shifted by $\pi$.  This has no effect on our conclusions.} and recalling that $\l$ is negative, we see that minimizing the potential requires maximizing $\cos 3\d_S$.  There are three equivalent solutions;
\begin{equation}
\d_S=0,~\frac23\pi,~\frac43\pi.
\end{equation}
Looking back at (\ref{gmass}) we see that the gaugino masses are real for each of these vacua.  This follows from the fact that our two phases were reduced to one by the equations of motion, and the $R$-symmetry rendered the remaining phase unphysical.  Had we neglected the gravitino, we still would have been able to rotate away our phase, but we will see that this freedom vanishes upon considering the Higgs sector.

We should mention that our three degenerate vacua pose a domain wall problem \cite{DWP}.  We will address the breaking of this $Z_3$ symmetry later.

\section{Generating $\mu$ and $B\mu$}

The Higgs mass terms can be generated by the same mechanism as above.  We introduce another singlet superfield $S'$ and replace the messenger with the Higgs fields, $H$ and ${\wt H}$.
\begin{equation}
W= \frac{\l'}{3}S'^3 + k'S'H{\wt H}+m_{3/2}M_P^2\label{WSp},\qquad K=|S'|^2-\frac{h'}{\Lambda^2}|S'|^4
\end{equation}
Just as before, we assume that $S'$ has a tachyonic SUSY-breaking soft mass, $V_{soft}\supset-m^2_{S'}$, and we rotate $S'$ such that $\l'$ is real.  The only difference is that the mass scale in this sector, $m_{S'}$, must be lower than that of the messenger sector to get a realistic spectrum.  This can be achieved with a single source of breaking provided that $S'$ couples to the SUSY-breaking sector more weakly than $S$ does\footnote{This can easily be accomplished with the model discussed in the appendix.}.

Borrowing our results from above, we see that the $F$-term of $S'$ is
\begin{equation}
F_{S'}=-\l'\langle S'^\dagger\rangle^2.
\end{equation}
From the superpotential in (\ref{WSp}), we then see that the $\mu$ and $B$ terms are
\begin{equation}
\mu =k'\langle S'\rangle,\qquad B=\frac{k'F_{S'}}\mu=-\l'|S'|e^{-3i\d_{S'}}.
\end{equation}
Using a phase rotation of $H{\bar H}$, we can take $\mu$ to be real.  The phase of $B$ is determined by the equations of motion, which give $\d_{S'}=0,\frac23\pi$, and $\frac43\pi$.  This means that we can simultaneously choose the gravitino mass, the gaugino mass, and the Higgs mass parameters to be real,
\begin{equation}
\arg m_{3/2}=\arg m_i=\arg\mu=\arg B=0,
\end{equation}
so there is no significant CP violation from sources outside of the Standard Model.

\section{The (non-)issue of sector mixing}

So far, we have neglected any interaction between the $S$ and $S'$ sectors.  This is justified in the superpotential because such terms cannot be generated perturbatively.  Moreover, it is not hard to invent dynamical models in which a single sector flows in the IR to two disjoint sectors (for a recent example, see \cite{Green}).  Mixing in the K\"ahler potential, however, cannot be forbidden.  In general, we have
 \begin{equation}
 K=|S|^2+|S'|^2-\frac{h|S|^4+h'|S'|^4+h''|S|^2|S'|^2}{\Lambda^2}+...
 \end{equation}
 The leading order contributions to the scalar potential from K\"ahler mixing are
 \begin{eqnarray}
 V_{mix} &=& \left(W_S+\frac{K_SW}{M_P^2}\right)g^{S\bar S'}\left(W_{S'}+\frac{K_{S'}W}{M_P^2}\right)^\dagger+h.c. \\ \nonumber &=&
 \frac{h''}{\Lambda^2}\left(m_{3/2}^2|S|^2|S'|^2+m_{3/2}\left(\l'|S|^2S'^3 +\l|S'|^2S^3\right)+\l\l'(SS'^\dagger)^3+h.c.\right)
 \end{eqnarray}
We see from examining the above terms that this mixing will not affect the constraints derived above.  Only the last term above is interesting because the first is real and the others only reinforce the previous constraints.  The term proportional to $\mbox{Re}(S'^3S^{\dagger3})\sim\cos(3(\d_S-\d_{S'}))$ forces the relative phase\footnote{The sum of these phases never appears in the scalar potential. This is because it is the Nambu-Goldstone mode associated with the $R$-symmetry breaking.} of $S$ and $S'$ to be (up to a $\pi$ ambiguity as before)
\begin{equation}
\d_{S}-\d_{S'}= 0,~\frac23\pi, ~\frac43\pi.
\end{equation}
This is, of course, automatically satisfied, so our solutions remain valid and degenerate.  In fact, we see here that we could set $m_{3/2}$ to zero and arrive at the same result.  The K\"ahler mixing term imposes $\arg(m_i B^*)=0$, preserving CP.

\section{Breaking the Discrete Symmetry}

The leftover $Z_3\times Z_3$ symmetry discussed above is undesirable. There is a domain wall problem \cite{DWP} since the potential can be minimized by three different phases for both $\d_S$ and $\d_{S'}$.
To solve this problem we introduce two Planck-suppressed explicit breaking terms;
\begin{equation}\label{wbreak}
W_{\rm breaking} = \frac{\k}{M_P} S^4 + \frac{\k'}{M_P}S'^4.
\end{equation}
Since these terms have a large suppression factor, we do not reintroduce a CP problem, and our previous analysis is nearly unaffected.  The important difference is that the degeneracy of vacua is lifted.

Mixing of the primed and unprimed sectors is irrelevant at leading order so it suffices to consider the following additional terms in the potential.
\begin{equation}
\Delta V=\frac{2|\k|}{M_P}\left(m_{3/2}|S|^4\cos(\d_\k+4\d_S)+4\lambda|S|^5\cos(\d_k+\d_S)\right),
\end{equation}
where $\d_\kappa$ is the phase of $\kappa$. To see that the vacuum degeneracy is broken, we add $\Delta V$ to (\ref{SV}) and find the new minima.  Up to ${\cal O}(\e^2)$ the three solutions are now
\begin{align}
\d_S&=-\e_1\sin\d_\k,&&V=V_0(1+\e_2\cos\d_\k)\frac{}{}\nonumber\\
\d_S&=\frac{2\pi}3-\e_1\cos(\d_\k+\pi/6),&&V=V_0(1-\e_2\sin(\d_\k+\pi/6))\nonumber\\
\d_S&=\frac{4\pi}3+\e_1\cos(\d_\k-\pi/6),&&V=V_0(1+\e_2\sin(\d_\k-\pi/6))
\end{align}
where
\begin{equation}
V_0=4h\l m_{3/2}\frac{|S|^5}{\Lambda^2},\qquad\e_1=\frac{2\k\Lambda^2(m_{3/2}+\l|S|)}{9h\l m_{3/2}|S|M_P},\qquad \e_2=\frac{\k\Lambda^2(m_{3/2}+4\l|S|)}{2h\l m_{3/2}|S|M_P.}
\end{equation}
Clearly the discrete symmetry is broken\footnote{The regions where two of the vacua become degenerate may still have a domain wall problem. However, dangerously degenerate vacua will only occur for very narrow regions of the phase $\d_k$.}. Because both the gaugino masses and $B$ are real for $\d_S,\d_S'=0,\frac{2}{3} \pi, \frac{4}{3}\pi$, the CP violation induced by $\d_k$ will be Planck suppressed.

\section{Conclusions}
	
In this work we have presented a mechanism that addresses the $\mu/B\mu$-problem of gauge mediation without introducing a CP problem.  The only source of CP violation that is introduced is Planck suppressed.  Moreover radiative generation of CP violation is utterly negligible despite the rather high energy scales involved.

It would be interesting to understand what other sorts of hidden sectors and messenger sectors can be employed in this mechanism.  It may be possible to construct a model, for example, without a residual discrete symmetry.

It would also be interesting to explore the phenomenology of such models, which should be quite distinctive.  Though the model resembles the NMSSM \cite{NMSSM} there are several important differences.  First of all, one assumes in the NMSSM that the gravitino mass does not give significant CP violation; if it is ${\cal O}(1)$ GeV or larger, this can be difficult to justify.  Assuming a basic gauge-mediated supersymmetry breaking, the gaugino masses can be taken real along with all coupling constants in the Higgs sector of the NMSSM.\footnote{We are grateful to Masahiro Ibe for pointing this out.} Clearly our mechanism is completely different in that the phases of the gaugino masses and $B$ are dynamically eliminated.  Indeed, the relevant phase is nothing but a Nambu-Goldstone boson---a dynamical variable much like that of the Peccei-Quinn mechanism.  Therefore, a generic prediction of our mechanism is the presence of an $R$-axion.

Furthermore, the NMSSM requires additional structure such as an extra pair of quarks, $Q$ and $\wt Q$, and a superpotential $W= S'Q \wt Q$ to induce a sufficiently large $\mu$ term \cite{Murayama}. Since the mass of the extra quarks is ${\cal O}(1)$ TeV, this model can be distinguished from the present model at the LHC.

Having made messengers out of the Higgses, we must address the usual problems with coupling supersymmetry too directly to the Standard Model.  For example, the Yukawa interactions give a negative one-loop mass to the stops.  However, because we have taken $F_{S'}\ll F_S$ this term will be a subleading contribution to the stop mass.  Other one-loop contributions to the scalar masses are less important because they are suppressed by small Yukawa couplings.

A more detailed discussion of the phenomenology of Higgs messenger models is left for future work \cite{ESY}.

\section*{Acknowledgments}
This work was supported by World Premier International Research Center Initiative (WPI Initiative), MEXT, Japan.

\appendix

\section{Dynamical origins of the model}

In this appendix we describe a particular UV extension of our model based on \cite{itiy} in which the mass scales that we introduced are generated by strong dynamics.  This extension is neither unique nor original.  We mostly follow \cite{luty} and \cite{IFY}.  The reader should consult these references for a more extensive discussion.

Consider an $SU(2)$ gauge theory with four doublet chiral superfields $Q_i$.  The meson formed from the gauge-invariant bilinear of these fields, $V_{ij}=\e_{\a\b}Q_i^\a Q_j^\b$, transforms as the {\bf 6} of the flavor $SU(4)\simeq SO(6)$.  In the quantum theory one finds that $Pf(V)=\Lambda^4$, which reduces the global symmetry \cite{Seiberg:1994bz}.  For the choice, $V=\Lambda^2\diag(\sigma^2,\sigma^2)$, an $Sp(4)\simeq SO(5)$ remains.  This is the breaking pattern that we will consider.

To simplify the analysis, we map our meson to a vector of $SO(6)$, $M_A$ with $A=0,\dots,5$.  The vev considered above then simply maps to $M_A$ getting a vev, which obviously leaves an $SO(5)$ global symmetry.  Adding six gauge singlets, we can add a manifestly $SO(6)$ invariant superpotential, $W=\Lambda Z_AM_A$ (the dynamical scale appears here because the corresponding UV operator is marginal).  Solving the quantum constraint, $M_AM_A=\Lambda^2$, for $M_0$ gives
\begin{eqnarray}
W_{eff}&=&\Lambda(Z_0\sqrt{\Lambda^2-M_aM_a}+Z_aM_a)\\
&\approx&\Lambda^2Z_0-\frac12Z_0M_aM_a+\Lambda Z_aM_a,\qquad a=1,\dots,5.
\end{eqnarray}
This is the classic O'Raifeartaigh model with a single, dynamically generated mass scale.  In the vacuum, supersymmetry is broken by $-F_{Z_0}^\dagger=\Lambda^2$.  This splits the masses of the $M_a$.

Now we introduce a messenger sector with the superpotential,
\begin{equation}
W=\k_E SE\wt{E}+\frac{\l}{3}S^3+kS\Psi\wt{\Psi}\label{WSE}
\end{equation}
This is natural in the sense that there is no renormalizable term that can be added without breaking some symmetry.  We can couple this sector to the SUSY-breaking sector by gauging a $U(1)\simeq SO(2)$ subgroup of the unbroken $SO(5)$ and taking $E$ and $\wt{E}$ to be charged along with, say, $M_1$ and $M_2$.  If this is done, supersymmetry breaking will be communicated to $E$ and $\wt{E}$ at two loops as in ordinary gauge mediation. This will lead to non-zero scalar masses for $E$ and $\wt{E}$,
\begin{equation}
m_{E}^2=m_{\wt{E}}^2\simeq \frac{\alpha_m}{4\pi}\frac{\Lambda^2}{16\pi^2,}
\end{equation}
which we've expressed in terms of the fine structure constant, $\a_m$, of the mediating $U(1)$. $E$ and $\wt E$ now act as messengers for $S$, giving it mass, but here the mass is generated at one-loop and is tachyonic, $V_{soft}\supset -m_S^2|S|^2$, where
\begin{equation}
m_{S}^2=4\frac{|\k_E|^2}{16\pi^2}m_{E}^2\ln\frac{\Lambda}{m_E}
\end{equation}
Including this contribution to the effective potential, one finds a global minima at
\begin{equation}
\vev S=e^{i\d_S}\frac{m_S}{\sqrt{2}\l},\qquad F_S =- \l \langle S^\dagger\rangle^2,
\end{equation}
and all other vevs zero.  Finally, because $m_S^2$ is loop suppressed relative to $m_E^2$, we can integrate out $E$ and $\wt E$ to give
\begin{equation}
W=\frac{\l}{3}S^3+kS\Psi\wt{\Psi}
\end{equation}
where
\begin{equation}
\langle S\rangle=\langle S\rangle+\theta^2F_S.
\end{equation}
This is precisely the setup discussed in the main text.  Proceeding in the same way with $S'$ and $E'$, we can generate our second sector.

\end{document}